\documentclass[12pt]{article}

\begin{document}

\newcommand{\be}{\begin{equation}}
\newcommand{\ee}{\end{equation}}
\newcommand{\ba}{\begin{array}}
\newcommand{\ea}{\end{array}}
\newcommand{\bea}{\begin{eqnarray}}
\newcommand{\eea}{\end{eqnarray}}
\newcommand{\bma}{\begin{matrix}}
\newcommand{\ema}{\end{matrix}}
\newcommand{\nn}{\nonumber}
\newcommand{\te}{\tilde{e}}

\begin{titlepage}
\vspace*{0.2in}
\begin{center}
{\large\bf  Brane Vector Dynamics from Embedding Geometry}
\end{center}
\vspace{0.2in}

\begin{center}
{T.E. Clark}\footnote{e-mail address: clark@physics.purdue.edu}$~^a~,~${S.T. Love}\footnote{e-mail address: loves@physics.purdue.edu}$~^{a}~,~${Muneto Nitta}\footnote{e-mail address: nitta@phys-h.keio.ac.jp}$~^c~,~
~${T. ter Veldhuis}\footnote{e-mail address: terveldhuis@macalester.edu}$~^{b}~,~${Chi Xiong}\footnote{e-mail address: xiong@purdue.edu}$~^a$\\
\end{center}
\begin{center}
{{\bf a.}~\it Department of Physics~,\\
 Purdue University~,\\
 West Lafayette, IN 47907-2036, U.S.A.}\\
\end{center}
\begin{center}
{{\bf b.}~\it Department of Physics \& Astronomy~,\\
 Macalester College~,\\
 Saint Paul, MN 55105-1899, U.S.A.}
\end{center}
\begin{center}
{{\bf c.}~\it Department of Physics~,\\
 Keio University~,\\
 Hiyoshi, Yokohoma, Kanagawa, 223-8521, Japan}\\
\end{center}
~\\

\abstract{
A Kaluza-Klein decomposition of higher dimensional gravity is performed in the flexible brane world scenario and  the properties of the extra vectors resulting from this decomposition are explored. These vectors become massive due to a gravitational Higgs mechanism in which the brane oscillation Nambu-Goldstone bosons become the longitudinal component of the vector fields. The vector mass is found to be proportional to the exponential of the vacuum expectation value of the radion (dilaton) field and as such its magnitude is model dependent. Using the structure of the embedding geometry, the couplings of these vectors to the Standard Model, including those resulting from the extrinsic curvature, are deduced.  As an example, we show that for 5D space-time the geometry of the bulk-brane world, either intrinsic or extrinsic, only depends on 
the extra vector and the 4D graviton. The connection between the embedding geometry and coset construction by non-linear realization is also presented.
}

\end{titlepage}
\newpage

\section{Introduction}
\vspace{0.3in}

If our world is a four dimensional brane floating in a higher dimensional space-time, an important physical consequence
is that the brane can fluctuate into the extra dimension(s). As such, some higher dimensional symmetries, such as  translation(s) along the
extra dimension(s), will be spontaneously broken\cite{brane} and there will appear  
the corresponding Nambu-Goldstone bosons. In the flexible brane limit where the scale that sets the brane tension, $ F_X $, is much smaller than the $D$-dimensional Planck scale, $ M_D $, the Kaluza-Klein modes of higher dimensional
gravity decouple from the Standard Model particles. When the broken higher dimensional symmetries are realized locally, a gravitational Higgs mechanism ensues and these Nambu-Goldstone modes become the longitudinal components of massive vector fields\cite{brane, Gpbranes}. The phenomenology of these (brane) vector fields has recently been considered\cite{pheno} and contrasted with that resulting from including only the longitudinal (branon) modes\cite{Kugo}-\cite{DM}.

Within a Kaluza-Klein formalism\cite{KK}, these extra vectors originate from the off diagonal components of the higher dimensional metric. Using this decomposition, it will be established that the vector mass depends exponentially on the vacuum expectation value of the radion (dilaton) field which is the scalar component in this decomposition. As such the value of the vector mass is model dependent. In particular, it could be in the TeV range and thus may be accessible to the LHC. The coupling of these vectors to the Standard Model and to gravity can be
obtained either via the method of nonlinear realizations of the spontaneously broken symmetries of higher dimensional space-time, or by the
embedding geometry of the bulk-brane world. This paper addresses the latter approach.
 
Section 2 provides a decomposition of higher dimensional gravity as in Kaluza-Klein theories resulting in an expression for the  brane vector's mass. The embedding frames and the embedding conditions are introduced in section 3 along with their integrability conditions which are described by Gauss-Weingarten equations
and Gauss-Coddazi-Ricci equations respectively.  It is then shown how the intrinsic and extrinsic geometry depend on the graviton and the brane
vectors. In section 4,  the connections between the embedding geometry and non-linear realization method is established. Finally, conclusions are presented in section 5.


\section{ Decomposition of Metric and Brane Vectors}


\vspace{0.3in}

The Kaluza-Klein formalism\cite{KK} provides the decomposition of the gravitational metric in $d>4$ dimensions into its various spin components in $d=4$. In general, the $d=4$ fields will consist of the spin-2 graviton, spin-1 vector fields and scalar (radion or dilaton) degrees of freedom. Traditionally, most applications of this formalism have attempted to unify gravity with the Standard Model and as such have identified the vector fields with the gauge bosons of the Standard Model. However, this does not have to be the case and the vectors could correspond to novel degrees of freedom. In this paper, they are identified with the vectors which emerge in flexible brane world models as a consequence of the spontaneous breaking of local space translation symmetries.

We begin by considering the zero modes of the 5D Kaluza-Klein
metric tensor 
\footnote{The (4+N) dimensional space-time  metric tensor $ \eta_{AB} $ has signature 
$( -, +, +, +, \cdots, +)$. Curved indices are denoted $ M, N, ...$ for the (4+N) dimensional space-time and $ \mu, \nu $ for the 4D theory, while the local Lorentz indices are $ A, B, ... $ for (4+N)-dimensions  and $ a, b, ... $ for 4D respectively . Finally, the indices $i, j, ...$ 
label the co-volume. 
}
\be  \label{5DKKmetric}
G_{MN}(x) =  \rho^{-\frac{1}{3}} \left(\ba{c c} g_{\mu\nu} +  \rho A_{\mu} A_{\nu} & ~~~ \rho A_{\mu} \\ 
\rho A_{\nu} & ~~~\rho \ea \right) ~.
\ee
Compactified on a circle with radius $ r $, the 4D effective action is \cite{KK}
\bea \label{4Deff}
\nn
S_G  &=&  -\frac{1}{2 \kappa^2_5} \int\!d^4x dy ~~ e^{(5)} R^{(5)}   \\
     &=&   -\frac{1}{2 \kappa^2} \int\!d^4x ~~ e^{(4)} [R^{(4)} + \frac{1}{4}  \rho F_{\mu\nu} F^{\mu\nu}
+ \frac{1}{6 \rho^2} \partial^{\mu} \rho \partial_{\mu} \rho ]
\eea
where $F_{\mu\nu}= \partial_{\mu}A_{\nu} -\partial_{\nu}A_{\mu} $ and $\kappa_5, \kappa$ are the 5D and 4D gravitational constants respectively which are related via $\kappa^2_5 = 2 \pi r \kappa^2 $. 
The indices are raised by $g^{\mu\nu} $ which is the inverse of $g_{\mu\nu}$.
The presence of a  3-brane in the 5D bulk breaks the extra 5D translation and Lorentz symmetry spontaneously. The position of the brane is provided by the embedding function $ Y^M = Y^M (x^{\mu}) $ with $x^{\mu}$  
the coordinates on the brane. The brane action is of the Nambu-Goto form built from the induced metric tensor \cite{Nambu}
$h_{\mu\nu}=G_{MN} \partial_{\mu}Y^M 
\partial_{\nu} Y^N$ and given by
\be \label{NGaction}
S_{\textrm{\scriptsize{brane}}}=  F_X^4 \int\!d^4x \sqrt{\textrm{det}~h_{\mu\nu}} ~.
\ee
We employ the static gauge defined by $ Y^{\mu} (x) = x^{\mu}, Y^5 (x) = \phi(x)$.
The Nambu-Goldstone boson $ \phi $ describes the brane fluctuation
for a 5D space-time with non-dynamical gravity. When we consider a curved 5D space-time with dynamical gravity
as (\ref{5DKKmetric}) and compactify the 5D theory on a circle, an extra vector field appears in the induced metric $h_{\mu\nu}$  after the field $\phi$ is absorbed as the longitudinal component by $ A_{\mu} $ \cite{brane}-\cite{DM}.   
Defining $ X_{\mu} \equiv A_{\mu} +  \partial_{\mu} \phi $, the induced metric can be written as 
\be \label{induce_metric}
h_{\mu\nu}= \rho^{-\frac{1}{3}} g_{\mu\nu} + \rho^{\frac{2}{3}}( A_{\mu} +  \partial_{\mu} \phi ) ( A_{\nu} + \partial_{\nu} \phi )  
=\rho^{-\frac{1}{3}} (g_{\mu\nu} +   \rho X_{\mu} X_{\nu}) ~.
\ee
Note that the field strength is $F_{\mu\nu} = \partial_{\mu}A_{\nu} -\partial_{\nu}A_{\mu} =  \partial_{\mu}X_{\nu} -\partial_{\nu}X_{\mu} $ and hence
the kinetic term of $A_{\mu}$ simply becomes the kinetic term of $ X_{\mu} $. 
The global limit is restored by taking 
\be
g_{\mu\nu} = \eta_{\mu\nu}~, ~~A_{\mu} =0~, ~~\rho = 1 ~.
\ee
Plugging the induced metric (\ref{induce_metric}) into the brane action (\ref{NGaction}) yields 
\bea 
\nn
S_{\textrm{\scriptsize{brane}}} &=&   \int\!d^4x ~F_X^4 \rho^{-\frac{2}{3}} \sqrt{\textrm{det}g_{\mu\nu} } \sqrt{\textrm{det}~(\delta^{\tau}_{~\lambda}+  \rho  X^{\tau} X_{\lambda} )} \\
                                &=&   \int\!d^4x ~F_X^4 \rho^{-\frac{2}{3}} \sqrt{\textrm{det} g_{\mu\nu} } (1+  \kappa^2 \tilde{X}^{\lambda} \tilde{X}_{\lambda}  + {\cal{O}}(X^4) ) ~.
\eea
To canonically normalize the Maxwell term in eq. (\ref{4Deff}), we rescaled the vector field as $ X_{\mu} =\kappa \sqrt{\frac{2}{\rho} }  ~\tilde{X}_{\mu} $, while the dilaton kinetic term in eq. (\ref{4Deff}) is put into canonical form after the  redefinition,  $ \rho = e^{\mp \sigma} $. So doing, the resulting vector mass is then gleaned as  
\be
m_{\tilde{X}}^2  \sim ~e^{ \pm \frac{2}{3} <\sigma>}  \kappa^2  {F_X^4} ~.
\ee
If one takes $ <\rho> =1$, i.e. $ <\sigma> =0 $, the mass of $ \tilde{X}_{\mu} $ is found to be 
$ m^2_{\tilde{X}} \sim  \kappa^2 F_X^4  $. For $ F_X \sim TeV $, 
this mass is very small \cite{Kugo, DM}.  However, in general, the size of the dilaton vacuum value is model dependent and consequently so is the vector mass. Thus the ``scaling factor",  may exponentially increase or suppress the 
mass of the vector depending on the form of the metric tensor in the extra dimensional space-time. 

Now consider the more general case where there are  $ N >1 $ co-dimensions. 
The 3-brane is embedded in a $(4 + N)$ -dimensional  bulk space-time with topology $ M_4 \times B $ and  coordinates $ Y^M = ( x^{\mu}, y^i )$ , where the co-volume $ B $  is a compact manifold with an isometry. 
The $ (4 + N ) $ dimensional metric is 
\be  \label{NDmetric}
G_{MN} = \left(\ba{c c} g_{\mu\nu}(x) +  \rho(x) \gamma_{ij}(y)  A^{i}_{\mu} (x,y) A^{j}_{\nu}(x, y) & ~~~  \rho(x)  \gamma_{kj}(y) A^{j}_{\mu} (x, y) \\ 
 \rho(x) \gamma_{jl} (y) A^{j}_{\nu} (x, y) & ~~~\rho (x) \gamma_{kl} (y) \ea \right)
\ee 
where $A^{i}_{\mu} (x,y) \equiv \xi^i_{\alpha} (y) A^{\alpha}_{\mu} (x) $ and $\xi^i_{\alpha} (y)$ are Killing vectors 
for describing the isometry of the co-volume. If the co-volume $B$ is homogeneous and isotropic, 
then its maximal isometry group can have $ \frac{1}{2} N(N+1) = N + \frac{1}{2} N(N-1) $ Killing vectors.
The 4-dimensional brane breaks all the isometries except the ones that belong to the stability group.  
More precisely, we denote $ \xi_{\alpha} = ( \xi_i, ~\xi_a ) $ and there are $ N $ Killing vectors $ \xi_i , i=1, 2, ..., N  $ which 
correspond to $ N $ broken translations due to the existence of the brane, i.e. $ \xi_i = \partial_i $ and  $ \frac{1}{2} N(N-1) $
 Killing vectors $ \xi_a,~a=1, 2, ...,\frac{1}{2} N(N-1)$, which correspond to $ \frac{1}{2} N(N-1) $ 
unbroken generators. These Killing vectors $ \xi_a $ may form an SO(N) Lie algebra as the cases considered in \cite{Gpbranes}, i.e. $ \xi_{jk} = \frac{1}{2} ( y_k \partial_{j} - y_j \partial_{k} ) $. Then one can also decompose
$  A^{\alpha}_{\mu} (x) = (A^i_{\mu}(x), ~A^{jk}_{\mu} (x) ) $.
Since \cite{KK, Cho1975}
\be
R^{4+N} = R^4 + \frac{1}{4}  \rho \gamma_{ij} \xi^i_{\alpha} \xi^j_{\beta} F^{\alpha}_{\mu\nu}  F^{\beta\mu\nu}  
+ L_{\textrm{\scriptsize{scalar}}}
\ee
where the scalar term $L_{\textrm{\scriptsize{scalar}}}$ can be calculated from ref. \cite{Cho1975}. The 4D effective action is 
\be \label{NDeff}
S_G  =  -\frac{1}{2 \kappa^2} \int ~d^4 x ~\sqrt{g}~[ R^{4} \rho^{\frac{N}{2}}  +  \frac{1}{4} \rho^{\frac{N+2}{2}} F^{\alpha}_{\mu\nu} 
F^{\mu\nu}_{\alpha} + \tilde{L}_{\textrm{\scriptsize{scalar}}} ] 
\ee
where we have used that
\bea
\nn 
\sqrt{G} & = & \sqrt{\textrm{det}~G_{MN}} = \sqrt{\textrm{det}~g_{\mu\nu}} \sqrt { \textrm{det}~\gamma_{ij}} ~  \rho^{\frac{N}{2}} ~,\\
       \kappa_D^2 & = &  \kappa^2    \int_B ~ d^N y  \sqrt{\gamma}
\eea
with 
\be
     \frac{\int_B ~ d^N y  \sqrt{\gamma} \gamma_{ij} \xi^i_{\alpha} \xi^j_{\beta} } {\int_B ~ d^N y  \sqrt{\gamma}} = \delta_{\alpha\beta} ~.
\ee
Here $\tilde{L}_{\textrm{\scriptsize{scalar}}}$ is obtained from $L_{\textrm{\scriptsize{scalar}}}$ by integrating over the extra dimensions. The brane action has the Nambu-Goto form 
\bea \label{NDNG}
\nn
S_{\textrm{\scriptsize{brane}}} &=& F_X^4 \int~d^4 x ~ \sqrt{\textrm{det} h_{\mu\nu}} \\
                                    &=& F_X^4 \int~d^4 x ~ \sqrt{\textrm{det}  (g_{\mu\nu} +  \rho \gamma_{ij} X_{\mu}^i  X_{\nu}^j } )
\eea 
where $ X_{\mu}^i (x) = A_{\mu}^i (x) + \partial_{\mu} \phi^i (x) $.\footnote{In the brane action (\ref{NDNG}), any $ y^i$-coordinate dependence 
of the metric (\ref{NDmetric}) and $A^i_{\mu}(x,y)$ is eliminated using the embedding functions $ y^i = y^i (x) = y^i_0 + \xi^i_j(y_0) \phi^j(x)$, where $y^i_0$ is a particular position of the brane.}
Analogously to the 5D case,  we rescale the metric, $g_{\mu\nu}=\tilde{g}_{\mu\nu}  \tilde{\rho}^{~-\frac{N}{N+2}}$, the
vector field, $ X_{\mu} = \kappa \sqrt{\frac{2}{\tilde\rho} }  ~\tilde{X}_{\mu} $and scalar field $ \rho= \tilde{\rho}^{~\frac{2}{N+2}} $ so that the higher dimensional metric (\ref{NDmetric}) takes the form
\be  \label{sNDmetric}
G_{MN} = \tilde{\rho}^{~-\frac{N}{N+2}} \left(\ba{c c} \tilde{g}_{\mu\nu} +   \tilde{\rho} \gamma_{ij} A^{i}_{\mu} A^{j}_{\nu}
 & ~~~   \tilde{\rho} \gamma_{kj} A^{j}_{\mu}   \\ 
  \tilde{\rho} \gamma_{jl}  A^{j}_{\nu}  & ~~~ \tilde{\rho}  \gamma_{kl} \ea \right) ~.
\ee 
With these rescalings the 
Einstein-Hilbert and Yang-Mills terms in the 4D effective action (\ref{NDeff}) assume their canonical form.  As in 5D case, we take $\tilde\rho= e^{ \mp <\sigma>}$
and the brane action becomes 
\be
S_{\textrm{\scriptsize{brane}}} = e^{ \pm \frac{2N}{N+2} <\sigma>  } F_X^4  \int~d^4 x ~ \sqrt{\textrm{det}  ( \tilde{g}_{\mu\nu} + 2 \kappa^2 \gamma_{ij} \tilde{X}^i_{\mu}  \tilde{X}^j_{\nu}  } )
\ee
from which we extract the vector mass term $ e^{\pm \frac{2N}{N+2}  <\sigma> } \kappa^2 F_X^4 \tilde{X}^{i\mu}  \tilde{X}_{i \mu} $. 
Thus for any extra dimensional space-time, there can be an exponential enhancement (or suppression) for the vectors masses. 
Note that this exponential factor is reminiscent of that employed by the Randall-Sundrum model \cite{Randall}
in relating the weak scale to the Planck scale.  
The fact that the vacuum expectation value (vev) of the dilaton can control various coupling constants is well known  
in string theory where the vacuum expectation value of the dilaton is related to the string coupling.  

When the 4D effective theory is constructed using the method of non-linear realizations \cite{Gpbranes}, the vector kinetic terms and mass terms arise as completely independent invariant Lagrangian monomials with the mass parameters arbitrary.     
Consequently, we treat the masses of these vectors as free parameters to be constrained by experiment. The couplings of these massive vectors to gravity and the Standard Model fields  
will be addressed in the next section by applying the embedding geometry and deriving the Einstein equation on the brane.
Included in these interactions will be derivative couplings of $X_{\mu} $  to the Standard Model fields which are related to the extrinsic curvature of the brane.

\section{Couplings of Brane Vectors to Gravity and Matter} 

\vspace{0.3in}

\subsection{Embedding geometry and Einstein equation on brane}

In this section, the general couplings of $X_{\mu} $ to matter and gravity are deduced using the 
embedding geometry \cite{embed}
of the bulk-brane world scenario. This approach has been previously used \cite{SMS} -\cite{pwest}  in
brane scenarios and now we apply it to the case of brane vectors. Introducing the embedding frame, $  \te_{\mu}=Y^M_{~,~\mu} \partial_M, ~n_i =n_i^M  \partial_M  $, 
with $ n_i , i =1, 2, ... N $ the normal vectors to the brane, the embedding conditions 
\bea \label{EC}
G_{MN} \partial_{\mu} Y^M \partial_{\nu} Y^N &=& h_{\mu \nu} ~,\cr
G_{MN} \partial_{\mu} Y^M n^N_i &=& 0 ~,\cr
G_{MN} n^M_i n^N_j &=& \delta_{ij}
\eea
relate the higher dimensional metric and the induced metric on the brane as well as provide the orthogonality condition
of $\te_{\mu}$ and $ n_i $ and the normalization of $ n_i $. 
Defining $ \nabla_{\mu} \equiv \te_{\mu}^{~M} \nabla_M $, where $ \Gamma_{MN}^{K} $ are the higher dimensional 
Christoffel connections , 
the covariant derivatives of the embedding frame basis are given by the Gauss-Weingarten equations \cite{SMS} -\cite{embed}
\bea \label{moveframe}
\nn
\nabla_{\mu} \te_{\nu}  &=& \Gamma^{\lambda}_{\mu\nu} \te_{\lambda} - K^i_{\mu\nu} n_i ~,\\
\nabla_{\mu} n^i  &=& K^{i}_{\mu\nu} \te^{\nu} + B^{ij}_{\mu} n_j 
\eea
which introduce the extrinsic curvature $ K^i_{\mu\nu} $, the 4-dimensional connection $ \Gamma^{\lambda}_{\mu\nu} $, 
 and the twist potential $ B^{ij}_{\mu}  $. These can be expressed in terms of the embedding frame basis and 
their covariant derivatives, using the embedding conditions (\ref{EC}) as 
\be \label{coeff}
 K^i_{\mu\nu} = - n^i_M \nabla_{\mu}(Y^M_{~,\nu})~,~~
\Gamma^{\lambda}_{\mu\nu} = h^{\lambda\sigma} Y^M_{~,\sigma} G_{MN} \nabla_{\mu} (Y^N_{~,\nu})~,~~
B_{\mu ij} = n_{Mj} \nabla_{\mu} n^M_i .
\ee
Since the covariant differentiation $\nabla_M $ is torsion free, the extrinsic curvature is symmetric \cite{embed}, $K^i_{\mu\nu} =K^i_{\nu\mu}$. Further note that the twist potential vanishes in the case of co-dimension one, $N=1$. 
Using the Gauss-Weingarten equations, it is straightforward to deduce 
their integrability conditions, the Gauss-Codazzi-Ricci equations \cite{SMS} -\cite{embed}, which relate 
the higher dimensional Riemannian curvature tensor to the lower dimensional induced one, 
plus the extrinsic curvature and the twist potential as
\bea \label{Gauss-Coddazi}
\nn
 \hat{R}_{KLMN}~ \te^{K}_{~~\rho} \te^{L}_{~~\sigma} \te^{M}_{~~\mu} \te^{N}_{~~\nu} 
 &=& R_{\rho\sigma\mu\nu} + K^i_{\mu\rho} K_{\nu\sigma i} - K^i_{\mu\sigma} K_{\nu\rho i} ~,\\ 
\nn
\hat{R}_{KLMN} ~ n^{Ki} \te^{L}_{~~\sigma} \te^{M}_{~~\mu} \te^{N}_{~~\nu} 
 &=& \tilde{\nabla}_{\mu} K^i_{\nu\sigma} -\tilde{\nabla}_{\nu} K^i_{\mu\sigma} ~,\\
\hat{R}_{KLMN} ~ n^{Ki} n^{Lj} \te^{M}_{~~\mu} \te^{N}_{~~\nu} 
 &=& F_{\mu\nu}^{ij} + K^i_{\mu\tau} K_{\nu}^{\tau j}  - K^i_{\nu\tau} K_{\mu}^{\tau j}
\eea
where 
\bea 
\nn
R^{\lambda}_{\tau \mu \nu} &=& \partial_{\nu} \Gamma^{\lambda}_{\mu\tau} -\partial_{\mu} \Gamma^{\lambda}_{\nu\tau}
 + \Gamma^{\sigma}_{\mu\tau}  \Gamma^{\lambda}_{\nu\sigma} -\Gamma^{\sigma}_{\nu\tau}  \Gamma^{\lambda}_{\mu\sigma} ~,\\
\nn
F_{\mu\nu}^{ij} &=& \partial_{\mu} B^{ij}_{\nu} -\partial_{\nu} B^{ij}_{\mu} + B^{ik}_{\nu} B^j_{\mu k} - B^{ik}_{\mu} B^j_{\nu k} ~,\\
\tilde{\nabla}_{\mu} K^i_{\nu\sigma} &=& \nabla_{\mu} K^i_{\nu\sigma} - B^{ij}_{\mu} K_{\nu\sigma j} ~.
\eea
These are the basic ingredients of the embedding geometry and now we apply them to the study of brane vectors. 
For simplicity, we consider a 5-dimensional space-time so the bulk-brane world has co-dimension one and the twist potential $ B^{ij}_{\mu} $ vanishes. 
In this case, we can remove all the $i,j $ indices and set $ B^{ij}_{\mu}=0 $ in the embedding condition (\ref{EC}), 
the Gauss-Weingarten equations (\ref{moveframe}) and the expression of $ K_{\mu\nu} $ in (\ref{coeff}). 
The last equation of (\ref{Gauss-Coddazi}) becomes trivial and the first two equations are also simplified yielding, 
\bea \label{5D-Gauss-Coddazi}
\nn
 \hat{R}_{KLMN}~ \te^{K}_{~~\rho} \te^{L}_{~~\sigma} \te^{M}_{~~\mu} \te^{N}_{~~\nu} 
 &=& R_{\rho\sigma\mu\nu} + K_{\mu\rho} K_{\nu\sigma} - K_{\mu\sigma} K_{\nu\rho}   ~,\\ 
\hat{R}_{KLMN} ~ n^{K} \te^{L}_{~~\sigma} \te^{M}_{~~\mu} \te^{N}_{~~\nu} 
 &=&  \nabla_{\mu} K_{\nu\sigma}  -\nabla_{\nu} K_{\mu\sigma} ~.
\eea
Using the 5D Einstein equation $ \hat{R}_{MN} - \frac{1}{2} G_{MN} \hat{R} = - \kappa_5^2 \hat{T}_{MN}$  where $ \hat{T}_{MN}$ is the 5D stress energy tensor of matter sources and following  \cite{SMS},  the Einstein equations on the brane take the form
\bea \label{braneeq}
\nn
R_{\mu\nu} - \frac{1}{2} R h_{\mu\nu}+ E_{\mu\nu} + Q_{\mu\nu} &=& -\frac{2}{3} \kappa_5^2 [ \hat{T}_{MN} \te^{M}_{~~\mu}  \te^{N}_{~~\nu} 
+ (\hat{T}_{MN} n^M n^N - \frac{1}{4} \hat{T} ) h_{\mu\nu} ] ~,\\
\nabla_{\tau} K^{\tau}_{\mu} - \nabla_{\mu} K &=& \kappa_5^2 ~ n^M \te^{N}_{~~\mu} \hat{T}_{MN}
\eea
where
\bea
\nn
E_{\mu\nu} &=& \hat{C}_{KLMN} ~ n^{K} n^{M} \te^{L}_{~~\mu} \te^{N}_{~~\nu} ~,  ~~\hat{T} = G^{MN} \hat{T}_{MN}  ~,\\
Q_{\mu\nu} &=& (K_{\mu\nu} K - K_{\mu \tau} K^{\tau}_{\nu}) - \frac{1}{2} h_{\mu\nu} (K^2-K_{\sigma\tau} K^{\sigma\tau})
~ , ~~K = h^{\mu\nu} K_{\mu\nu} = \textrm{Tr}K
\eea
with $\hat{C}_{KLMN}$ the 5D Weyl tensor. 
We first address the case where the 5D space-time is  flat, 
\footnote{Strictly speaking the 5D space-time cannot be flat due to the presence of the brane as the matter source. However, we assume that it does not bend the 5D space-time much so the metric can be considered as an almost flat one. }
and work in static gauge defined as 
$ Y^{\mu} = x^{\mu}, ~Y^5 = \phi(x)$  so that
\bea \label{inphi}
\nn
&& \te^{\nu}_{~\mu} = \delta^{\nu}_{~\mu} ~~,~~~~\te^{5}_{~\mu}=\partial_{\mu} \phi~~, ~~~~ E_{\mu\nu}=0  ~,\\
\nn 
&& n_{\mu}=-\partial_{\mu} \phi /\sqrt{1+\partial_{\tau} \phi ~\partial^{\tau} \phi}~~,
~~~n_{5}=1/\sqrt{1+\partial_{\tau} \phi ~\partial^{\tau} \phi}   ~,\\  
\nn
&& h_{\mu\nu}=\eta_{\mu\nu} + \partial_{\mu} \phi~ \partial_{\nu} \phi~~, ~~  
h^{\mu\nu}=\eta^{\mu\nu} - \partial^{\mu} \phi ~\partial^{\nu} \phi / (1+\partial_{\tau} \phi ~\partial^{\tau} \phi ) ~,\\
\nn
&& K_{\mu\nu} = -\partial_{\mu} \partial_{\nu} \phi / \sqrt{1+\partial_{\tau} \phi ~\partial^{\tau} \phi}~~ ~,\\
&& R_{\rho\sigma\mu\nu} = (\partial_{\mu} \partial_{\sigma} \phi ~\partial_{\nu} \partial_{\rho} \phi -
    \partial_{\mu} \partial_{\rho} \phi ~\partial_{\nu} \partial_{\sigma} \phi ) / (1+\partial_{\tau} \phi ~\partial^{\tau} \phi ) ~.
\eea 
It follows  that  the only physical degree of freedom is the Nambu-Goldstone boson, $ \phi $, which describes 
the fluctuation of the brane.  
Equations (\ref{braneeq}) are consistency equations for $\phi$ and its derivatives. Since the extra 
dimensional translation is spontaneously broken, the dynamics of the Nambu-Goldstone field $\phi$ can be secured using the
conservation of the broken symmetry current $ \partial_{\mu} \hat{T}^{\mu 5} = F_{\phi}^2 \partial^2 \phi + ... = 0 $.
Alternatively, the field equation follows from a minimization the trace of extrinsic curvature as shown in \cite{Sorokin}. This is equivalent to the p-brane
equations of motion which one obtains from the Nambu-Goto action. It corresponds geometrically to the minimal volume obtained from the 
embedding of the corresponding world volume into higher dimensional space-time.  In this case, if the brane is the  only
matter source in 5D spacetime, then the vanishing condition of the trace of the extrinsic curvature $K = h^{\mu\nu} K_{\mu\nu} =0$ leads to
\be
\partial^2 \phi = \frac{\partial^{\mu} \phi ~\partial^{\nu} \phi ~\partial_{\mu} \partial_{\nu}\phi }{ 1+\partial_{\tau} \phi ~\partial^{\tau} \phi }
\ee
which is recognized as the same equation of motion of $\phi$ as that obtained from the Nambu-Goto action \cite{Nambu, Sorokin}. Also the extrinsic curvature can be related to the rigidity of strings or branes \cite{Polyakov}. 
In general it is difficult to solve these equations and, moreover, the form of the stress energy tensor $ \hat{T}_{MN} $ must be specified. However, the equations for $\phi$ can be converted to an action which includes the leading couplings like $\partial_{\mu}\phi~\partial_{\nu}\phi ~T^{\mu\nu}_{SM} $ plus other higher order derivative terms.

\subsection{Brane vector and its couplings}
\vspace{0.2in}

Next consider  a curved 5D space-time with the general Kaluza-Klein metric of eq. (\ref{5DKKmetric}). The Gauss-Coddazi equations and the induced Einstein equation now 
become more complicated producing a set of differential equations for the spin-2 (4D graviton 
$g_{\mu\nu}$), spin-1 (4D vector $ A_{\mu}$ ) and spin-0 (4D dilaton $\rho$). As discussed in refs. \cite{brane, Kugo, DM}, the Higgs mechanism is operational. Naively, one simply replaces $ \partial_{\mu}\phi \rightarrow \partial_{\mu}\phi + A_{\mu} \rightarrow X_{\mu} $ in eq. (\ref{braneeq}). Some care is required, however, since $ \partial_{\mu} \partial_{\nu} \phi $ has the ambiguity of being replaced by either $\partial_{\mu} X_{\nu} $ or  $\partial_{\nu} X_{\mu}$. Moreover, there are also terms dependent on the field strength $ F_{\mu\nu} $. To resolve any ambiguity, 
one must figure out the relation between $n_{\mu} $ and $X_{\mu}$. To do so, the 5D Kaluza-Klein metric 
(\ref{5DKKmetric}) is used to calculate the Christoffel connections and extrinsic curvature which are shown to depend only on $ X_\mu$ and $g_{\mu\nu} $.  

Consider the transformation laws of the various fields. 
A bulk vector field $ V^M $  transforms under a general coordinate transformation as  
\be
\delta_{\epsilon} V^M = \epsilon^K \partial_K V^M - V^K \partial_K \epsilon^M  
\ee
where $ \epsilon^M = ( \epsilon^{\mu}(x, y), \epsilon^5 (x, y) ) $.
As in the usual 5D Kaluza-Klein theories, we take $ \epsilon^{\mu} = 0 $, and $ \epsilon^{5} = \epsilon (x)$. This  corresponds to
 a particular 5D general coordinate transformation (or a gauge transformation for $A_{\mu} $)
\be \label{kktransf}
 Y^{' \mu} = Y^{ \mu}(x)~, ~~ Y^{'5} = Y^5 - \epsilon(x)
\ee
so that $ \phi(x) =Y^5(x) $. In addition, the zero mode fields $g_{\mu\nu}(x), ~A_{\mu}(x),~ \rho(x),  ~V^{\mu}(x), ~V^5(x)$ transform as
\bea 
\nn
\delta_{\epsilon} \phi ~~&=&  - \epsilon (x)~, ~~~~~~~~~~~~~\delta_{\epsilon} g_{\mu\nu} =~0 ~,\\
\nn
\delta_{\epsilon} A_{\mu} &=& \partial_{\mu} \epsilon (x)~, ~~~~~~~~~~~~\delta_{\epsilon}\rho~~~=~ 0 ~,\\
\delta_{\epsilon} V^{5}   & =& - V^{\mu} \partial_{\mu}\epsilon (x) ~, ~~~~~~\delta_{\epsilon} V^{\mu}  ~=~ 0 ~.
\eea
It is easy to see that $X_{\mu} \equiv A_{\mu} + \partial_{\mu} \phi, ~ \hat{V^5} \equiv A_{\mu} V^{\mu} + V^5, ~\hat{V^{\mu}}  
\equiv V^{\mu},  ~\rho $ and $g_{\mu\nu} $ are all invariant under the transformation (\ref{kktransf}).  From eq.(\ref{induce_metric}), it follows that the induced metric $h_{\mu\nu} = \rho^{-\frac{1}{3}} (g_{\mu\nu} +   \rho X_{\mu} X_{\nu}) $ and all intrinsic geometric quantities on the brane, such as the Christoffel connection, Riemannian tensor, Ricci tensor and Ricci scalar are invariant as well. The normal vector and the extrinsic curvature can then be computed. 
 
In Section 2, we discussed the role that $\rho$ plays in modifying the mass of brane vectors. Here, to simplify the calculation, we set $ \rho =\kappa=1 $ and decompose the 5D metric as 
\bea  \label{Guassdecomp}
\nn
G_{MN} &=& \left( \ba{c c}  g_{\mu\nu} +  A_{\mu} A_{\nu} &  A_{\mu} \\  A_{\nu} &  1 \ea \right) \\
&=& S^T \hat{G} S 
\eea 
with $ S = \left( \ba{c c}  \delta_{~~\nu}^{\tau}  &  0 \\  A_{\nu}  & 1  \ea \right) $ and 
$ \hat{G} = \left( \ba{c c}  g_{\rho \tau}  &  0 \\  0 & 1  \ea \right) $. The matrix $ S $ when acting on a bulk vector shifts only the fifth component so that, for example,  
\be
 \hat{n}^{M} \equiv S^M_{~K} n^K = (n^{\mu}, ~n^5 + A_{\nu} n^{\nu}) ~.
\ee
This provides an invariant form under the transformation (\ref{kktransf}) provided one takes $V^M = n^M $. Acting
on $ \te^M_{~, \mu} = Y^M_{~, \mu} $ in the static gauge gives  
\be
\hat{Y}^M_{~, \mu} \equiv S^M_{~K} Y^K_{~, \mu} = ( \delta^{\nu}_{~ \mu} ~,  ~ A_{\mu} + \partial_{\mu} \phi ) = 
( \delta^{\nu}_{~ \mu} ~,  ~ X_{\mu} ) ~.
\ee
The embedding conditions can be written in this ``shifted" frame as 
\be 
h_{\mu\nu} = g_{\mu\nu} + X_{\mu} X_{\nu}~, ~~~\hat{n}^{\mu} = - X^{\mu} \hat{n}^5~, ~~~g_{\mu\nu} \hat{n}^{\mu}\hat{n}^{\nu} + (\hat{n}^5)^2 =1
\ee
which can be readily solved yielding
\be \label{nmu5}
\hat{n}^{\mu} = \frac{- X^{\mu}}{ \sqrt{1+ X^{\mu} X_{\mu}}}~, ~~~~ \hat{n}^5 = \frac{1}{ \sqrt{1+ X^{\mu} X_{\mu}}}
\ee
where $ X^{\mu} = g^{\mu\nu} X_{\nu} $.  
To compute $  \Gamma^{\lambda}_{\mu\nu} $ and $ K_{\mu\nu} $, only the first equation (Gauss equation) of (\ref{moveframe}) needs to be solved. Multiplying by the matrix $ S^L_{~~M} $ on both sides gives
\bea
\nn
S^L_{~~M} \nabla_{\mu} Y^M_{~, \nu} &=& \Gamma^{\lambda}_{\mu\nu}  S^L_{~~M} Y^M_{~, \lambda} 
- K_{\mu\nu} S^L_{~~M} n^M  \\
&=& \Gamma^{\lambda}_{\mu\nu}  \hat{Y}^L_{~, \lambda} 
- K_{\mu\nu}  \hat{n}^L ~.
\eea
To compute the left hand side, we use the 5D metric $ G_{MN} $ to compute the connections 

\bea \label{5Dconnection}
\nn
\bar{\Gamma}^{\lambda}_{\mu\nu} &=& \tilde{\Gamma}^{\lambda}_{\mu\nu} - \frac{1}{2} g^{\lambda\rho} ( A_{\mu} F_{\rho\nu} +A_{\nu} F_{\rho\mu}) ~,\\
\nn
\bar{\Gamma}^{5}_{\mu\nu} &=&  \frac{1}{2} A^{\rho} ( A_{\mu} F_{\rho\nu} +A_{\nu} F_{\rho\mu}) 
+ \frac{1}{2} (  \tilde{\nabla}_{\nu} A_{\mu}+ \tilde{\nabla}_{\mu} A_{\nu}     ) ~,\\
\bar{\Gamma}^{\lambda}_{5 \mu} &=& -\frac{1}{2} g^{\lambda\rho}  F_{\rho\mu}, ~~~~
\bar{\Gamma}^{5}_{5 \mu} = \frac{1}{2} A^{\rho}  F_{\rho\mu}
\eea
where $F_{\mu\nu} = A_{\nu, \mu} - A_{\mu, \nu} =  X_{\nu, \mu} - X_{\mu, \nu}$, 
 and $   \tilde{\Gamma}^{\lambda}_{\mu\nu} $ is built from $g_{\mu\nu}$. It follows that
\bea
S^L_{~~M} \nabla_{\mu} Y^M_{~, \nu} &=& \bigg \{  \ba {c c}  
\tilde{\Gamma}^{\lambda}_{\mu\nu}  - \frac{1}{2} g^{\lambda \rho} ( F_{\rho\mu} X_{\nu} +F_{\rho\nu} X_{\mu} )  & ; ~~~{}_{L=\lambda}  \\
\frac{1}{2}  (X_{\mu, \nu} + X_{\nu, \mu} )  & ; ~~~{}_{L=5}  \ea ~.
\eea
Now  $\Gamma_{\tau\mu\nu} $ and $K_{\mu\nu}$ are computed as
\bea \label{K}
\nn
\Gamma_{\tau\mu\nu} &=& \tilde{\Gamma}_{\tau\mu\nu}  - \frac{1}{2} ( F_{\tau\mu} X_{\nu} +F_{\tau\nu} X_{\mu} ) +  
\frac{1}{2}  X_{\tau} (X_{\mu, \nu} + X_{\nu, \mu} ) ~,\\
K_{\mu\nu} &=& -\frac{1}{ 2 \sqrt{1+ X^{\mu} X_{\mu}}} [ X^{\rho} (F_{\rho\mu} X_{\nu} +F_{\rho\nu} X_{\mu} ) + (\tilde{\nabla}_{\nu} X_{\mu}+ \tilde{\nabla}_{\mu} X_{\nu} ) ]
\eea
where $\tilde{\nabla}_{\nu} X_{\mu} \equiv X_{\mu, \nu} - \tilde{\Gamma}^{\lambda}_{\mu\nu} X_{\lambda} $. The Christoffel connection $\Gamma_{\tau\mu\nu} $ coincides with the result computed directly from the induced metric. 
When taking the flat 5D space-time limit, the extrinsic curvature $K_{\mu\nu} $ reduces to the previously obtained result (\ref{inphi}).  This expression is a generalization of the so-called ADM formulation of gravity \cite{ADM}. Since we did not include the higher Kaluza-Klein modes,  a term like $\partial_y g_{\mu\nu} $ vanishes in $K_{\mu\nu}$. 
Note that besides $ g_{\mu\nu} $, the only other field dependence in the induced metric, intrinsic curvature,
extrinsic curvature etc. occurs through the combination $ X_{\mu} = A_{\mu} + \partial_{\mu} \phi $. Moreover, as noted previously, 
the Maxwell term $F_{\mu\nu} F^{\mu\nu}$ for $A_{\mu}$, which was obtained from the decomposition of 5D Einstein-Hilbert term $ \hat{R}^{(5)} $, does not change when replacing  $A_{\mu}$ by $X_{\mu}$ so
\be
\hat{R}^{(5)} = R^{(4)} (g) + \frac{1}{4} F_{\mu\nu}F^{\mu\nu} , ~~F_{\mu\nu} = X_{\nu, \mu} - X_{\mu, \nu} ~.
\ee
On the other hand, one can derive a different decomposition of $ \hat{R}^{(5)} $ using eq. (\ref{5D-Gauss-Coddazi}) as
\be \label{5D_ADM}
\hat{R}^{(5)} = R^{(4)} (h) + \textrm{Tr}K^2 - (\textrm{Tr}K)^2 + 2\nabla_M ( n^M \nabla_N n^N-n^N \nabla_N n^M) 
\ee
where $\textrm{Tr}K^2=K_{\mu\nu} K^{\mu\nu}, \textrm{Tr}K = K_{\mu\nu} h^{\mu\nu}$ and $h_{\mu\nu} = g_{\mu\nu} + X_{\mu} X_{\nu}$ 
(Note that $R^{(4)} (h) $ is calculated from the induced metric $h_{\mu\nu} )$.  Integrating and noting that we
only include the zero modes, one obtains
\be \label{RKF}
\int~d^4x \sqrt{g} ~[ R^{(4)} (h) - R^{(4)} (g) + \textrm{Tr}K^2 - (\textrm{Tr}K)^2 ] =  \int~d^4 x \sqrt{g}~\frac{1}{4} F^2 
\ee
which is a simple relation among the scalar curvature $ R^{(4)} $, the extrinsic curvature terms $ \textrm{Tr}K^2,  (\textrm{Tr}K)^2 $
and the Maxwell term $ F^2$. It is easy to check this identity at the order of ${\cal O} (X^2) $ and this provides an alternative 
way to build the Maxwell term, which will be discussed from the point of view of a 4D non-linear realization in the next section.

The expression for the normal vector can be used to extract the couplings of $X_{\mu}$ to gravity and the Standard Model. 
Notice that the right hand side of the first equation in (\ref{braneeq}) contains the term $ \hat{T}_{MN} n^M n^N h_{\mu\nu}  $. Since both $ \hat{T}_{MN} $ and $n^M$ do not contain $ h_{\mu\nu}  $ explicitly, this term in the Einstein equation must correspond to an action term 
\be \label{tmunu}
\sqrt{\textrm{det}h_{\mu\nu}} \hat{T}_{MN} n^M n^N = \sqrt{h}~ ( T^{SM}_{\mu\nu} n^{\mu} n^{\nu} + \cdots ) 
\ee
where we assume that the stress-energy tensor of the Standard model $T^{SM}_{\mu\nu}$ is included in $ \hat{T}_{MN} $ 
as in ref. \cite{SMS} and the ellipsis represents other components of  the $\hat{T}_{MN}$ term. Using the form for 
$ \hat{n}^{\mu} $ (c.f. eq. (\ref{nmu5})) 
\be 
n^{\mu} = \hat{n}^{\mu} =  \frac{- X^{\mu}}{ \sqrt{1+ X^{\mu} X_{\mu}}} = - X^{\mu} + {\cal O} (X^2) ~.
\ee
Plugging into (\ref{tmunu}) and expanding in the power series of $ X_{\mu} $, 
one readily extracts the lowest order couplings of $ X_{\mu} $ to the Standard Model as
i.e. 
\be
 \sqrt{h}~ T^{SM}_{\mu\nu} n^{\mu} n^{\nu} \sim  \sqrt{g}~ X^{\mu} X^{\nu} T^{SM}_{\mu\nu} + {\cal O} (X^4) ~.
\ee
This is the non-derivative coupling of brane vector $X_{\mu} $ to the Standard Model fields.

Next consider the derivative couplings to the Standard Model. 
The $ X_{\mu} $ field strength  
\be
F_{\mu\nu} =  X_{\nu, \mu} - X_{\mu, \nu} = \tilde{\nabla}_{\mu} X_{\nu} - \tilde{\nabla}_{\nu} X_{\mu}
\ee
and (c.f. eq. (\ref{K})) 
\be
K_{\mu\nu} = - \frac{1}{2} ( \tilde{\nabla}_{\mu} X_{\nu} + \tilde{\nabla}_{\nu} X_{\mu} ) + {\cal O} ( X^2) 
\ee
contain the anti-symmetric and symmetric pieces of $\partial_{\mu} X_{\nu}$
respectively. Since both $F_{\mu\nu}$ and $K_{\mu\nu}$ are invariant under the transformation (\ref{kktransf}), so is their product 
\bea \label{KF}
\nn
F_{\mu \rho} K^{\rho}_{~\nu} &=& F_{\mu \rho} g^{\rho \tau } K_{\tau\nu}  \\
                                                &=& \tilde{\nabla}_{[\nu} X_{\rho}  \tilde{\nabla}^{\rho} X_{\mu]}  + 
\frac{1}{2} (\tilde{\nabla}^{\rho} X_{\mu}  \tilde{\nabla}_{\rho} X_{\nu} -  \tilde{\nabla}_{\mu} X^{\rho}  \tilde{\nabla}_{\nu} X_{\rho} ) +  {\cal O} ( X^3)  ~.   
\eea
Notice that the first term of last line of (\ref{KF}) is anti-symmetric in $\mu, \nu $ while the second term is symmetric. Since $ X_{\mu} $ is a singlet under the Standard Model gauge group, the above combination couples invariantly to the $ SU(3) \times SU(2) \times U(1) $ singlet antisymmetric hypercharge field strength $B_{\mu \nu} $ and its dual $ \tilde{B}_{\mu\nu} $ as
\be \label{BKF}
(\kappa_1 B^{\mu\nu} + \kappa_2 \tilde{B}^{\mu\nu} ) F_{\mu \rho} K^{\rho}_{~\nu} =  (\kappa_1 B^{\mu\nu} + \kappa_2 \tilde{B}^{\mu\nu} ) \tilde{\nabla}_{[\nu} X_{\rho}  \tilde{\nabla}^{\rho} X_{\mu]} +  {\cal O} ( X^3) ~.   
\ee
Here $\kappa_1, \kappa_2 $ are dimensionless parameters.
This interaction has the same dimension as the $ X^{\mu} X^{\nu} T_{\mu\nu}$ terms. 
In addition, $X_{\mu}$ has invariant couplings to the Standard Model scalar doublet bilinear, $\varphi^{\dagger} \varphi$, which can contribute to the decay rate of the Standard Model Higgs boson \cite{pheno}.
Combining terms (and taking the $g_{\mu\nu} = \eta_{\mu\nu} $ limit ) yields the effective action 
\bea \label{action}
\nn
S_{4D~eff} = \int\!d^4x [~L_{SM} &-& \frac{1}{4} F_{\mu\nu} F^{\mu\nu} +\frac{1}{2} m^2_X X^{\mu} X_{\mu} ~]  \\
\nn
&+& \frac{m^2_X}{2F_X^4} [~T^{\mu\nu}_{SM} X_{\mu} X_{\nu} + 
(\kappa_1 B^{\mu\nu} + \kappa_2 \tilde{B}^{\mu\nu} ) F_{\mu \rho} K^{\rho}_{~\nu}  \\
&+& (\lambda_1 K_{\mu\nu} K^{\mu\nu}+\lambda_2 F_{\mu\nu}F^{\mu\nu} + \lambda_3  F_{\mu\nu}\tilde{F}^{\mu\nu} ) (\varphi^{\dagger}  \varphi - \frac{v^2}{2}) ~] ~.
\eea
The phenomenology based on the effective action (\ref{action}) has been 
studied in ref.\cite{pheno}. For $N\ge 2$ isotropic codimensions, there 
is an $SO(N)$ symmetry associated with the isometry of the co-dimensional 
space. Under this symmetry, the $X^\mu_i$ vectors transform 
non-trivially while all Standard Model fields are $SO(N)$ singlets. As 
such, invariant couplings of the $X^\mu_i$ vectors must occur in pairs 
and the vectors are stable. For the $N=1$ case, the stability is 
insured  provided there is a discrete reflection symmetry in the extra 
dimension under which the vector is odd, $ X_{\mu} \longrightarrow 
-X_{\mu} $.

\vspace{0.2in}

\section{Connection to Coset Method} 

\vspace{0.2in}
In the last two sections, we deduced the couplings of the brane vectors from the bulk-brane world point of view 
using the embedding geometry. So doing, we constructed a four dimensional effective action detailing their interactions with gravity and matter fields. The non-linear realization, or coset, method, provides another approach to four dimensional effective theories. Previously, the relation between
the embedding geometry and the coset method has been considered \cite{Kugo, Bandos, Sorokin, pwest} 
for the case of embedding a  hypersurface  into a  flat higher dimensional space-time.
Here we consider the properties of a $(p+1)$-dimensional hypersurface
embedded into a curved $D$-dimensional space-time with dynamical gravity. In fact, since the brane vectors arise from the off-diagonal components of the higher dimensional metric, their presence requires that 
the higher dimensional space-time  be curved. Previously\cite{Gpbranes} we showed  how to generalize the coset formulation 
in order to include the gravitational fields. In this section, we examine the connection between the two formalisms.

\subsection{Embedding Geometry in Moving Frames}

Thus far, we have employed the coordinate bases ($  \te_{\mu} = Y^M_{~,\mu} \partial_M, ~ n_i = n_i^M  \partial_M  $) to 
describe the embedding geometry. However, in order to make 
connection with the coset approach and construct the Maurer-Cartan forms, we need Lie algebra valued matrices. 
To achieve this correspondence, the embedding  conditions can be recast\cite{brane, Kugo, Bandos, Sorokin, pwest} as  
\bea \label{embed}
\nn
E^A_{~M} ~ \partial_{\mu} Y^M ~(U^{-1})^a_{~A} &=& e^a_{~\mu} ~,\\
E^A_{~M} ~ \partial_{\mu}  Y^M ~(U^{-1})^i_{~A} &=& e^i_{~\mu} ~=~0
\eea
where $ E^A_{~M} $ is the higher dimensional vielbein and $ U^B_{~A} = ( U_{~~a}^B, ~ U_{~~i}^B ) $ are $SO(1, D-1)$ matrices whose inverse is $(U^{-1})_{~B}^{A} \equiv \eta^{AC}\eta_{BD} U^D_{~C}  \equiv  ( (U^{-1})_{~B}^{a},~ (U^{-1})_{~B}^{i} ) $.
Conditions (\ref{embed}) show that $E^A_{~M} ~ \partial_{\mu} Y^M $ is not the induced vielbein on the brane  and one has to perform a Lorentz rotation $ U^B_{~A} $ to ensure that the induced vielbein and the 
induced metric satisfy $ h_{\mu\nu} = e^a_{~\mu} e^b_{~\nu} \eta_{ab} $\cite{brane, Kugo}. 
First we briefly review the properties of the 
$ U^B_{~A} $ matrices, which have been discussed in detail in \cite{Kugo, Bandos, Sorokin}.  By definition, they satisfy 
\be \label{UOC}
U^A_{~C} U^B_{~D}  \eta^{CD} =   \eta^{AB} , ~~~  U^A_{~C} U^B_{~D}  \eta_{AB} =   \eta_{CD}
\ee
which are invariant under the independent  left and right  $SO(1, D-1)_L \times SO(1, D-1)_R $ transformations 
\be 
U^A_{~B} \longrightarrow \tilde{U}^A_{~B} = ( \Lambda_L )^A_{~C} U^C_{~D} (\Lambda^{-1}_R)^D_{~B} ~.
\ee
Equations (\ref{embed}) are invariant under the transformations $SO(1, p)  \times SO(N) $ and thus the group $SO(1, D-1)_L$ is
broken down to $SO(1, p)  \times SO(N) $, while the $SO(1, D-1)_R$ group is unbroken. Therefore these  $U^A_{~B}$ matrices
parametrize a coset manifold $ \frac{SO(1, D-1)}{SO(1, p) \times SO(N) }$ and the embedding is Lorentz covariant due to the unbroken $SO(1, D-1)_R$ symmetry.

Next we build connections between these two frames (actually the connections among the coordinate basis, non-coordinate basis induced on the brane and the bulk geometry ).  The first equation of (\ref{embed}) shows how $ e^a_{~\mu} $ is related to 
$ E^A_{~M} \partial_{\mu} Y^M $. If we define the normal vectors as 
\be \label{new-n}
n^i_M \equiv E^A_{~M} (U^{-1})^i_{~A} 
\ee
then the second equation of (\ref{embed}) is just the orthogonality condition $ n^i_M \partial_{\mu}  Y^M = 0$. 
With definition (\ref{new-n}), it is easy to see that $ G_{MN} n_i^M n_j^N = \eta_{ij} = \delta_{ij} $ which is the normalization condition of 
vectors $ n_i $. This shows that our definition for $ n_i $ satisfies both orthogonality and normalization conditions for normal vectors.  
Note, however, that $ n_i $ is fixed only up to an $SO(N)$ rotation. 
The relations can be summarized as 
\bea \label{ncb}
\nn
e_a &=& U_{~a}^{A} E_A = U_{~a}^{A} E_A^{~M} \partial_M = e_a^{~\mu} \partial_{\mu} Y^M \partial_{M}  ~,\\
n_i  &=& U_{~i}^{A} E_A = U_{~i}^{A} E_A^{~M} \partial_M =  n_i^M \partial_{M} ~.
\eea 
Also recall the coordinate basis used in the previous section 
\bea \label{cb}
\nn
\tilde{e}_{\mu} &=& \partial_{\mu} Y^M \partial_{M}  ~,\\
n_i  &=&  n_i^M \partial_{M} ~.
\eea
Now we write the Gauss-Weingarten equations in the non-coordinate basis, with new coefficients $ \omega^{c}_{ab} , 
~K^i_{ab}$ and $ B^{ij}_{b}$ to be determined, as
\bea \label{GW} 
\nn
\nabla_{e_a} e_b  &=& \omega^{c}_{ab} e_{c} - K^i_{ab} n_i  ~,\\
\nabla_{e_a} n^i    &=& K^{ic}_{a} e_{c} + B^{ij}_{a} n_j ~.
\eea
After some straightforward, albeit lengthy, calculation, we find the Gauss-Coddazi-Ricci equations in the non-coordinate basis are given by
\bea \label{GCR}
\nn
 \hat{R}_{ABCD}~ U_{~a}^{A}  U_{~b}^{B} U_{~c}^{C} U_{~d}^{D}
 &=& R_{abcd} + K^i_{ac} K_{bdi} - K^i_{bc} K_{adi}   ~,\\ 
\nn
\hat{R}_{ABCD} ~ U_{~i}^{A}  U_{~b}^{B} U_{~c}^{C} U_{~d}^{D}
 &=& \tilde{\nabla}_{c} K_{bdi} - \tilde{\nabla}_{d} K_{bci}  ~,\\
\hat{R}_{AB}^{~~~CD} ~ U_{~a}^{A}  U_{~b}^{B} U_{C}^{i} U_{D}^{j}
 &=& F_{ab}^{ij} + K^i_{ac} K_{b}^{cj}  - K^i_{bc} K_{a}^{cj}
\eea
where $U_{C}^{j} = (U^{-1})_{C}^{j} $. In obtaining this result, we have employed the torsion free condition $\omega^c_{ab} -\omega^c_{ba} =\mathcal{C}_{ab}^c  $. 
Note that the curvatures built from the spin connection and twist potential  contain the anholonomy coefficient $ \mathcal{C}_{ab}^c  $ terms as 
\bea
\nn
R^c_{dab} &=& e_b \omega^c_{ad} - e_a \omega^c_{bd} + \omega^c_{be} \omega^e_{ad} -\omega^c_{ae} \omega^e_{bd} +\mathcal{C}_{ab}^e \omega^c_{ed} ~,\\
\nn
F_{ab}^{ij} &=& e_{a} B^{ij}_{b} -e_{b} B^{ij}_{a} + B^{ik}_{b} B^j_{a k} - B^{ik}_{a} B^j_{b k} - \mathcal{C}_{ab}^c B^{ij}_c  ~,\\ 
\tilde{\nabla}_{a} K^i_{bc} &=& e_a K^i_{bc} -\omega^d_{ac} K^i_{bd} -\omega^d_{ab} K^i_{dc}- B^{ij}_{a} K_{bcj} ~.
\eea
Using equations (\ref{ncb}), the components of eq. (\ref{GW}) can be written in terms of the $ U $ matrices as
\bea \label{GWU} 
\nn
\hat{\nabla}_{a} U_{~b}^{A}  &=& \omega^{c}_{ab} U_{~c}^{A} - K^i_{ab} U_{~i}^{A}  ~,\\
\hat{\nabla}_{a} U_{~i}^{A}    &=& K_{ai}^b U_{~b}^{A} + B^{~j}_{ai} U_{~j}^{A} 
\eea
where $ \hat{\nabla}_{a} \equiv U_{~a}^{A} ( E_A + \Omega_A) $ and 
$ \hat{\nabla}_{a} U_{~m}^{A}  = U_{~a}^{D} (E_D^{~M} \partial_M U_{~m}^{A} + \Omega_{DC}^A U_{~m}^{C} ) $, for $ m = (b, i) $. 
Here $ \Omega^A_{BC} $ is the higher dimensional spin connection. 
From equation (\ref{UOC}), we obtain the expressions for the extrinsic curvature, spin connection and twist potential respectively as
\bea \label{newcoeff}
\nn
 K^i_{ab} &=& - (U^{-1})^i_{~A} \hat{\nabla}_{a} U_{~b}^{A}  ~,\\
\nn
\omega^{c}_{ab} &=& + (U^{-1})^c_{~A} \hat{\nabla}_{a} U_{~b}^{A} ~,\\
B_{a i}^{j} &=& + (U^{-1})^j_{~A} \hat{\nabla}_{a} U_{~i}^{A} ~.
\eea
One may recognize the $ U^{-1} \hat{\nabla} U $ pattern in  the above equations and consider them as the components of 
a covariant Cartan form (This will be shown manifestly in the next section through the coset approach)
\be
{(U^{-1} \hat{\nabla} U )}^{A}_{~B} = \left( \ba{c c}  \omega^a_{~b}   & K^a_{~j}  \\ 
-K^i_{~b} & B^i_{~j}  \ea \right) 
\ee
with the one-forms  $ K^i_{~b} \equiv  e^a K^i_{ab} ,  ~B^i_{~j}  \equiv e^a B_{a j}^{i}  , 
~ \omega_{~b}^{c} \equiv  e^a \omega^{c}_{ab} $.
It follows that the  $ K^i_{ab} $  and $ K^i_{\mu\nu} $, $B^{ij}_a$ and $ B^{ij}_{\mu}$,  $\omega^{c}_{ab} $ and $ \Gamma^{\lambda}_{\mu\nu} $ are simply related as
\bea \label{cntn}
\nn
 K^i_{ab}  &=& e_a^{~\mu} e_b^{~\nu} K^i_{\mu\nu}  ~,\\
\nn
 B^{ij}_a    &=& e_a^{~\mu} B^{ij}_{\mu} ~,\\
\omega^{c}_{ab}  &=& e^c_{~\mu}  e_a^{~\nu} (\partial_{\nu} e_b^{~\mu} + \Gamma^{\mu}_{\nu\lambda} e_b^{~\lambda})
= e^c_{~\mu}  e_a^{~\nu}  \nabla_{\nu} e_b^{~\mu} ~.
\eea
Note the last equation in (\ref{cntn}) is just the usual relation between spin connection and Christoffel symbol. 
 As mentioned earlier, the coset manifold of $ \frac{SO(1, D)}{SO(1, p) \times SO(N) } $ is  parametrized by the matrices $ U^A_{~B}  = ( e^{ i v_{ai} M^{ai} } )^A_{~B} $ where $ M^{ai} $ are the broken Lorentz generators. Following \cite{Kugo}
one obtains
\bea \label{UAB}
(U^{-1})^A_{~B} &=& \left( \ba {c c} \textrm{cos} \sqrt{v \hat{v} } & \displaystyle{\frac{\textrm{sin} \sqrt{v \hat{v} } }{\sqrt{v \hat{v} } } } v \\
- \hat{v} \displaystyle{\frac{\textrm{sin}\sqrt{v \hat{v} } }{\sqrt{v \hat{v} } } } & \textrm{cos} \sqrt{ \hat{v}  v } \ea \right)
\eea
where $ v = v^a_{~i}, \hat{v} = v^j_{~b} = \delta^{ji} \eta_{ba} v^a_{~i} $. The embedding condition 
$ E^A_{~M} \partial_{\mu}  Y^M (U^{-1})^i_{~A} = 0 $  imposes $ 4 N $ constraints which are the same as the number
of  the Nambu-Goldstone fields $ v^a_{~i} $.  Therefore these constraints completely fix $ v^a_{~i} $ in terms of
$ E^A_{~M} ~ \partial_{\mu}  Y^M $, though in general it is difficult to solve these constraints. Here we only need
the explicit expression for the induced vielbein $ e^a_{~\mu} $ ( for calculation details see \cite{Kugo} )
\be
e^a_{~\mu} = e_{\parallel~\lambda}^a ( 1+ M)^{\frac{1}{2}~\lambda}_{~~~\mu} ~, 
~~~M = (e^T_{\parallel} \eta e_{\parallel} )^{-1} e^T_{\perp} \delta e_{\perp}
\ee
where $  e_{\parallel~\mu}^a =  E^a_{~M} ~ \partial_{\mu}  Y^M, ~
e_{\perp~\mu}^i =  E^i_{~M} \partial_{\mu}  Y^M  $ and $ \eta = \eta_{ab}, \delta=\delta_{ij} $. 
Taking the Kaluza-Klein vielbein as (indices with bars are the co-volume world ones)
\be  \label{5Dvielbein}
E^A_{~M} = \left(\ba{c c}  \mathcal{E}^a_{~\mu} &  0 \\ 
\mathcal{E}^i_{~\bar{k}} \xi^{\bar{k}}_{\alpha} A^{\alpha}_{\mu}  &  \mathcal{E}^i_{~\bar{j}} \ea \right)
\ee
yields the induced vielbein on the brane 
\be \label{inde}
e^a_{~\mu} = \mathcal{E}^a_{~\lambda} ( \delta^{\lambda}_{~\mu} + X^{\lambda \bar{i} } X_{\mu \bar{i}} )^{\frac{1}{2}} 
\ee
which depends only on $  \mathcal{E}^a_{~\lambda} $ and $ X^i_{\mu} $  
(note that $ \mathcal{E}^a_{~\mu} \mathcal{E}^b_{~\nu} \eta_{ab} = g_{\mu\nu} $).
Therefore one can start with the embedding frame (coordinate basis) and compute $K^i_{\mu\nu}, B^{ij}_{\mu} $
in that frame and then use the induced vielbein to convert them to the ones in the non-coordinate frame, and finally obtain
the covariant Cartan forms which may be used to build an invariant effective action in 4D space-time. 

\subsection{ Connecting with the Coset Approach}

\vspace{0.2in}

Previously, we presented \cite{Gpbranes} a detailed construction of the $X$ vector coupling to gravity and the Standard Model using coset methods.  In that case,   
 a $p$-brane is embedded in  $ D $ dimensional space-time resulting in the spontaneous breakdown of the symmetry
group from $ ISO(1, D-1) $ to $ ISO(1, p) \times SO(N), N= D-p-1$. The $ISO(1, D-1) $ generators
$( M_{AB}, ~P_C ) $ are decomposed into those of the stability group $ SO(1, p) \times SO(N) $ generators
$( M_{ab}, M_{ij} )$, the broken Lorentz generators $ M_{ai} $, the broken translation generator $ P_i $
and the unbroken translation generators $ P_a $.  The coset element is taken to be 
\be  \label{coset_element}
\Omega (x) = e^{i x^a P_a } e^{i \phi^i (x) P_i } e^{ i v^{ai} (x) M_{ai}} ~.
\ee
A connection term which includes gravitational fields is then added to the Maurer-Cartan form so that
\be \label{MCform}
\omega = \Omega^{-1} \nabla \Omega \equiv \Omega^{-1} (d + i \hat{E}) \Omega 
\ee
transforms analogously to the way it did in the global case
\be
\omega^\prime(x^\prime) = h(x)\omega (x) h^{-1}(x) +h(x)dh^{-1}(x) ~,
\ee
with the stability group element $h(x) \in SO(1, p) \times SO(N)$. 

To ascertain the meaning of the embedding condition in the coset method, consider the general one form
\be 
 \omega = {\cal G} ^{-1} (d + i \hat{E} )  {\cal G} 
\ee
with $ {\cal G}  = {\cal P} {\cal U} , 
~{\cal P} \equiv e^{i Y^A P_A} ,~ {\cal U} \equiv ~e^{ \frac{i}{2} v^{BC} M_{BC} } $ and  
$ \hat{E} \equiv dY^M (\hat{E}^A_{~M} P_A - \frac{1}{2} \Omega_M^{~BC} M_{BC} ) $. Thus 
\bea \label{dform}
\nn
\omega   &=&  {\cal U} ^{-1}  {\cal P} ^{-1} ( d + i \hat{E} )  {\cal P}  {\cal U}   \\
                                      &=& dY^M [ i  E^A_{~M}   {\cal U} ^{-1} P_A  {\cal U}  +    {\cal U} ^{-1} \partial_M {\cal U}  
                                               - \frac{i}{2}  {\cal U} ^{-1} \Omega_M^{~BC} M_{BC}  {\cal U}  ]
\eea
where $ E^A_{~M}  \equiv  \hat{E}^A_{~M} + \delta^A_{~M} -  \Omega_M^{~AB} Y_B $ is the shifted vielbein 
\cite{Gpbranes}. The above one form has been 
decomposed according to the generators of $SO(1, D-1)$, i.e. the first term of the last line in (\ref{dform}) takes values on
 $P_A$, the second and third terms take values on $ M_{BC}$.  Now consider the  embedding of the brane whose position is described by
the embedding function $ Y^M = Y^M (x^{\mu}) $ as before.
Notice that the first term can be written as
\bea \label{pterm}  \nn
i ~dY^M   E^A_{~M}   {\cal U} ^{-1} P_A  {\cal U} &=& i~dx^{\mu} \partial_{\mu} Y^M  E^A_{~M} {\cal U} ^{-1} P_A  {\cal U} \\
&=&  i~ dx^{\mu} \partial_{\mu} Y^M  E^A_{~M}  L^{~B}_A P_B
\eea
where $ L^{~B}_A$ forms a vector representation of the $SO(1, D-1)$ Lorentz group.  Imposing the embedding condition as in (\ref{embed})
\be
E^A_{~M} ~ \partial_{\mu} Y^M ~L^{~a}_{A} = e^a_{~\mu}~, ~~
E^A_{~M} ~ \partial_{\mu}  Y^M ~L^{~i}_{A} = e^i_{~\mu} = 0
\ee
restricts the $SO(1, D-1)$ Lorentz matrices $ L^{~B}_{A} $ to be coset elements of 
$ \frac{SO(1, D-1)}{SO(1, p) \times SO(N) }$ as we have shown below equation (\ref{UOC}). In other words,
we may parametrize ${\cal U} = e^{ i v^{ai} M_{ai} } $, instead of a general  $SO(1, D-1)$ Lorentz
group element  $e^{ \frac{i}{2} v^{BC} M_{BC} } $, from the outset. Consequently $ {\cal G} =
e^{i Y^A P_A } e^{ i v^{ai} M_{ai} } $ is exactly the coset element  in (\ref{coset_element}), which after taking
the static gauge, i.e. $ Y^a = x^a, ~Y^i = \phi^i $, takes the form
\be
e^{i Y^A P_A } e^{ i v^{ai} M_{ai} }  = e^{i x^a P_a } e^{i \phi^i (x) P_i } e^{ i v^{ai} (x) M_{ai}} 
\ee
while $ {\cal G} ^{-1} (d + i \hat{E} )  {\cal G} $ becomes the covariant Maurer-Cartan one form.

As in the flat background case of
 higher dimensional space-time \cite{Bandos, pwest}, it follows using the Poincare algebra $ISO(1, D-1)$ commutators that
\bea  \label{L-PL}
\nn
 {\cal U} ^{-1} P_a  {\cal U} &=& (\textrm{cos} \sqrt{v \hat{v} })_a^{~b} P_b  - 
(\displaystyle{\frac{\textrm{sin} \sqrt{v \hat{v} } }{\sqrt{v \hat{v} } } } )_a^{~c} v_c^{~j} P_j  ~,\\
 {\cal U} ^{-1} P_i  {\cal U}  &=& \hat{v}_i^{~c} (\displaystyle{\frac{\textrm{sin}\sqrt{v \hat{v} } }{\sqrt{v \hat{v} } } } )_c^{~b} P_b
+ (\textrm{cos} \sqrt{ \hat{v}  v } )_i^{~j} P_j ~.
\eea
Comparing with (\ref{UAB}), it is easy to see that  by taking the vector 
representations for the broken Lorentz generators $ M_{ai} $ that $ (L^T)^{B}_{~A} = (U^{-1})^{B}_{~A} $.  
The angular momentum generator piece in the decomposition of (\ref{dform}) is then computed as
\be \label{omega}
\omega = dY^M [{\cal U} ^{-1} \partial_M {\cal U}  - \frac{i}{2}  {\cal U} ^{-1} \Omega_M^{~BC} M_{BC}  {\cal U}  ]
\ee
with $~{\cal U} = e^{ i v^{ai} M_{ai} } $.  Once again, taking the matrix representation for all  $SO(1, D-1)$ Lorentz generators as 
$ (M^{AB})^C_{~D} = i \eta^{CE} ( \delta^A_E  \delta^B_D - \delta^A_D  \delta^A_E) $, then  
$ ({\cal U})^{B}_{~A} = U^{B}_{~A} $, and eq. (\ref{omega}) becomes
\bea
\nn
\omega  &=&  dY^M [ U^{-1} \partial_M  U + U^{-1} \Omega_{M} U  ] \\
\nn
&=& dx^{\mu}  ~\partial_{\mu} Y^M E^B_{~M}   [(U^{-1})^a_{~B} (U^{-1}  \hat{\nabla}_a  U) +(U^{-1})^i_{~B} ( U^{-1}  \hat{\nabla}_i  U ) ]
\\
&=& dx^{\mu}  ~e^a_{~\mu} (U^{-1}  \hat{\nabla}_a  U)
\eea
with $\hat{\nabla}_a = U^A_{~a} (E_A + \Omega_A )$ (c.f. below equation (\ref{GWU})).  
Here  the embedding condition has been used in obtaining the last identity.  
Writing $ \omega = \frac{1}{2} \omega_{AB} M^{AB} $ and using the vector representation for $M^{AB}$, 
the identification 
\be 
\omega^A_{~B} =\left( \ba{c c}  \omega^a_{~b}   & \omega^a_{~j}  \\ 
\omega^i_{~b} & \omega^i_{~j}  \ea \right) 
= \left( \ba{c c}  \omega^a_{~b}   & K^a_{~j}  \\ 
-K^i_{~b} & B^i_{~j}  \ea \right)  
\ee
is secured. Thus, using the embedding condition (\ref{embed}), it is established that the covariant Maurer-Cartan 1-form
components, the induced vielbein, the induced spin connection, the
extrinsic curvature and the twist potential, all have geometrical meanings.  The coset approach and the embedding geometry construction yield identical results.  As an example, consider the 5D space-time case where the twist potential  vanishes, $ B^{ij}_{\mu} =0$, while  $ K^i_{\mu\nu} $ and $ \Gamma^{\lambda}_{\mu\nu} $ are given 
in eq. (\ref{K}), and the induced vielbein is simply 
$e^a_{~\mu} = \mathcal{E}^a_{~\lambda} ( \delta^{\lambda}_{~\mu} + X^{\lambda} X_{\mu} )^{\frac{1}{2}} $ (c.f. eq. (\ref{inde})).   Hence all the components of the covariant Maurer-Cartan 1-form can be explicitly expressed in terms of 
gravitational vielbein $ \mathcal{E}^a_{~\mu}, X_{\mu}$  and their derivatives.

We end this section by considering other embedding conditions.  
So far we have analyzed eq.(\ref{EC}) using the metric $ G_{MN} $
and eq.(\ref{embed}) using the vielbein $E^A_{~M}$. Note that the metric tensor can be expressed
in two different forms, called the K-K form and the ADM form, as
\bea  \label{KK_ADM}
\nn
G_{MN} &=& \left( \ba{c c}  g_{\mu\nu} +  g_{\bar{m} \bar{n}} A^{\bar{m}}_{\mu} A^{\bar{n}}_{\nu} & 
 g_{\bar{m} \bar{j}} A^{\bar{m}}_{\mu} \\  g_{\bar{i} \bar{m}} A^{\bar{m}}_{\nu} &  g_{\bar{i} \bar{j}} \ea \right) ~~~~~~~~~~~~~~~~ \textrm{\scriptsize{\underline{K-K}}} \\
              &=& \left( \ba{c c}  h_{\mu\nu}  &    N_{\bar{m} \bar{j}} N^{\bar{m}}_{\mu} \\  
 N_{\bar{i} \bar{m}} N^{\bar{m}}_{\nu} &  N_{\bar{i} \bar{j}} + N_{\bar{i} \bar{m}} N_{\bar{j} \bar{n}} h^{\lambda\tau} N^{\bar{m}}_{\lambda} N^{\bar{n}}_{\tau}  \ea \right) .~~~\textrm{\scriptsize{\underline{ADM}}} 
\eea 
These two forms come from the different decompositions of the metric tensor, i.e. 
$ G_{MN} = \eta_{AB} \mathcal{E}^A_{~M}\mathcal{E}^B_{~N} = \eta_{AB} e^A_{~M} e^B_{~N}$ where the K-K vielbein
$\mathcal{E}^A_{~M}$ and the ADM vielbein $e^A_{~M}$ will be given below. 
The fields $ ( g_{\mu\nu}, A^{\bar{m}}_{\mu} ,  g_{\bar{i} \bar{j}} ) $ and $ (h_{\mu\nu}, N^{\bar{m}}_{\mu} ,  N_{\bar{i} \bar{j}} ) $
are related as
\be
h_{\mu\nu}=g_{\mu\nu} +  g_{\bar{m} \bar{n}} A^{\bar{m}}_{\mu} A^{\bar{n}}_{\nu}~, ~~(N^{-1})^{ \bar{m} \bar{n}} = g^{\bar{m} \bar{n}} + A^{\lambda \bar{m}} A_{\lambda}^{\bar{n}}~,~~ N^{\bar{m}}_{\mu} = (N^{-1})^{ \bar{m} \bar{n}}  A_{\mu \bar{n}}  ~.
\ee
Now consider embedding the brane into the bulk spacetime. 
The original embedding condition (\ref{embed}) corresponds to the ADM form
\be \label{embed_1}
E^A_{~M} ~ \partial_{\mu} Y^M ~(U_1^{-1})^B_{~A} =  \bigg \{ \ba{l}e^a_{~\mu} \\  0 \ea
\iff
e^A_{~M} = \left(\ba{c c}  e^a_{~\mu} &   \eta^{ab}e^{~\lambda}_{b} N_{\lambda}^{\bar{k}} N_{\bar{k}\bar{j}} \\ 
0  &  e^i_{~\bar{j}} \ea \right) ~.
\\
\ee
That is, an arbitrary higher dimensional vielbein $E^A_{~M}$ is projected by $\partial_{\mu} Y^M$ and rotated by $ U_1^{-1} $
into the first column of the ADM vielbein. Alternatively, a rotation by $U_2^{-1} $ produces the K-K vielbein as
\be \label{embed_2}
E^A_{~M} ~ \partial_{\mu} Y^M ~(U_2^{-1})^B_{~A} = \bigg \{ \ba{l} \mathcal{E}^a_{~\mu} \\
\mathcal{E}^i_{~\bar{k}} X^{\bar{k}}_{~\mu} \ea 
\iff
\mathcal{E}^A_{~M} = \left(\ba{c c}  \mathcal{E}^a_{~\mu} &  0 \\ 
\mathcal{E}^i_{~\bar{k}} A^{\bar{k}}_{\mu}  &  \mathcal{E}^i_{~\bar{j}} \ea \right) ~.
\\
\ee
This corresponds to the condition used in the coset construction \cite{Gpbranes}.
Note that $X^{\bar{i}}_{~\mu}$ and $A^{\bar{i}}_{~\mu}$ coincide in the unitary gauge for the 
Nambu-Goldstone fields, i. e. $ \phi^i = 0 $.
Since these two vielbeins are related by an $SO(1, D-1)$ matrix $ T $ as $\mathcal{E}^A_{~M} = T^A_{~B} e^B_{~M} $, one can multiply (\ref{embed_1}) by the
matrix $ T^A_{~B} $, giving $ U_2 = T U_1 $. The explicit expression for $T^A_{~B}$ is 
\be
T^A_{~B} = \left( \ba {l l} 
\mathcal{E}^a_{~\mu} e^{\mu}_{~b} & 
-g_{\bar{m}\bar{n}} h^{\mu\nu} \mathcal{E}^a_{~\mu} A_{\nu}^{\bar{n}}e_j^{~\bar{m}} \\
\mathcal{E}^i_{~\bar{k}} A_{\lambda}^{\bar{k}} e_b^{~\lambda} & 
\mathcal{E}^i_{~\bar{k}} (\delta^{\bar{k}}_{\bar{m}} - N_{\bar{m}\bar{n}} A^{\lambda\bar{k}} A^{\bar{n}}_{\lambda} )e_j^{~\bar{m}} \ea \right) .
\ee
The embedding condition (\ref{embed_2}) also requires that the expression of normal vectors be modified. 
Formally, $n^i_M =  E^A_{~M} (U_2^{-1})^i_{~B} T^B_{~A} $ or more precisely 
\be \label{kk-n}
n^i_M = ({N^{-\frac{1}{2}}})^i_{j}~ [E^A_{~M} (U_2^{-1})^j_{~A} -  \tilde{e}^{\nu}_{M} \mathcal{E}^j_{~\bar{k}} X^{\bar{k}}_{~\nu} ] \ee
where $\tilde{e}^{\nu}_{M} = G_{MN} h^{\mu\nu} \partial_{\mu} Y^N$ is given in section 2. Note that the normal vectors are
determined up to $SO(N)$ rotations. Both conditions (\ref{embed_1}) and (\ref{embed_2}), with 
corresponding expressions (\ref{new-n}) and (\ref{kk-n}) for the normal vectors, lead to the same embedding condition
(\ref{EC}). The condition (\ref{embed_1}) is related to the embedding geometry more closely, while (\ref{embed_2}) splits the vielbein directly into the graviton and the brane vectors and is more convenient for phenomenological applications.

\section{Conclusions} 
\vspace{0.2in}

It has been shown that Kaluza-Klein gravity in higher dimensional space-time, combined with the brane world scenario, 
leads to extra vectors which couple to 4D gravity and the Standard Model.  The off diagonal components of the higher dimensional metric  become massive vector fields $ X^i_{\mu} $ as a consequence of the gravitational Higgs mechanism. 
As an example, a 4 dimensional brane embedded in a 5D space-time was considered and intrinsic and extrinsic geometrical objects, such as the induced
metric, connections, extrinsic curvature and so on were calculated. All these quantities depend on the 4D graviton and the vector $ X_{\mu} $. It follows that $ X_{\mu} $ is a salient dynamical degree of freedom for describing the
fluctuation of the brane.  Both
non-derivative and derivative couplings between $X_{\mu}$ and the Standard Model fields were studied and a four dimensional 
effective action was constructed from the higher dimensional theories and embedding geometry.
Finally the relation 
between the embedding and the coset approach was clarified by comparing the covariant Maurer-Cartan forms.

\section*{Acknowledgments}

\vspace{0.3in}
The work of TEC, STL and CX was supported in part by the U.S. Department of Energy under grant DE-FG02-91ER40681 (Task B).  
The work of M.N. is supported in part by Grant-in-Aid for Scientific
Research (No.~20740141) from the Ministry of Education, Culture, Sports, Science and Technology-Japan.
The work of TtV was supported in part by a Cottrell Award from the Research Corporation and by the NSF under grant 
PHY-0758073. CX would like to thank Martin Kruczenski for the discussions on string theory.

\end{document}